\documentclass[pre,aps,superscriptaddress,preprint,showpacs]{revtex4}

\usepackage{epsfig}
\usepackage{latexsym}
\usepackage{graphicx}
\usepackage{amssymb}

\begin{document}

\title{Phase diagram in inflating balloon}

\author{Fanlong Meng}
\affiliation{State Key Laboratory of Theoretical Physics, Institute of Theoretical Physics, Chinese Academy of Sciences, Beijing 100190, China}
\affiliation{Kavli Institute for Theoretical Physics China, Beijing 100190, China}
\affiliation{Center of Soft Matter Physics and its Applications, Beihang University, Beijing, China}

\author{Masao Doi}
\email{masao.doi@buaa.edu.cn}
\affiliation{Center of Soft Matter Physics and its Applications, Beihang University, Beijing, China}

\author{Zhongcan Ouyang}
\affiliation{State Key Laboratory of Theoretical Physics, Institute of Theoretical Physics, Chinese Academy of Sciences, Beijing 100190, China}
\affiliation{Kavli Institute for Theoretical Physics China, Beijing 100190, China}
\affiliation{Center for Advanced Study, Tsinghua University, Beijing 100084, China}

\date{\today}

\begin{abstract}
Volume transition of an inflating cylindrical balloon
made of rubber under external force $F$ is studied based on the non-linear elastic theory for rubber. The pressure difference $\Delta P$ between inside and outside of the balloon is calculated as a function of the volume $V$ of the fluid inside the balloon. It is shown that  if $F$ is smaller than a critical value $F_c$, the inflation of the balloon takes place with coexistence of strongly inflated region and weakly inflated region, while if $F$ is larger than $F_c$, the inflation takes place uniformly. It is shown that the critical force $F_c$ depends on the material parameter of the
rubber, and that the discontinuous transition for finite stretching force can take place only when the rubber has enough stretchability.
\end{abstract}

\maketitle

\begin{figure}[h]
\begin{center}
\includegraphics{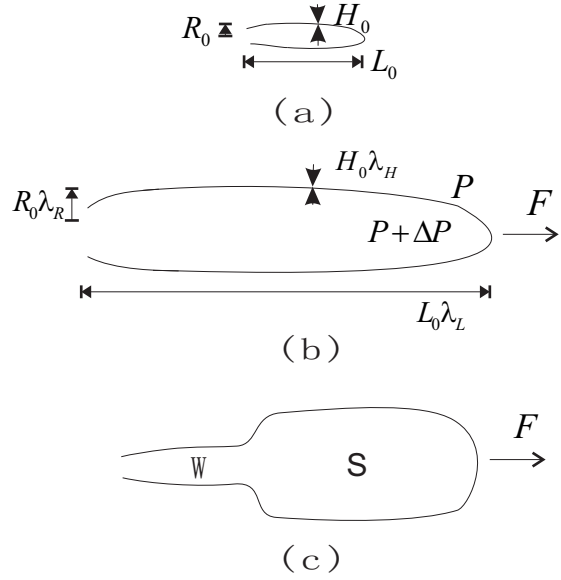}
\caption{
Inflation of a cylindrical balloon.  (a) denotes the initial state, (b)denotes the uniform inflation and (c) denotes the biphasic inflation when it is inflated by pressure $\Delta P$ subjected to an external stretching force $F$. In (b) the elongation in the circumferential and the axial directions are respectively denoted by  $\lambda_R = R/R_0$ and $\lambda_L= L/L_0$. (c) denotes the state of coexistence of strongly inflated region and weakly inflated region. S, W indicates strongly inflated and weakly inflated part of balloon, respectively.}
\label{fg1}
\end{center}
\end{figure}

When a balloon made of rubber is inflated, the inflation often takes place non-uniformly. For example, when a cylindrical balloon is inflated, the balloon first inflates uniformly as shown in  Fig.~\ref{fg1}~(b) , but at some point, a strongly inflated part appears ( Fig.~\ref{fg1}~(c)) and further inflation takes place by the growth of this part. Eventually the inflation becomes uniform again (Fig.~\ref{fg1}~(b)).

This phenomenon is a kind of volume transition, where the equilibrium volume of a system changes discontinuously when external parameters are changed continuously.
Such phenomenon is known for many systems,
typically the gas-liquid transition in simple liquids, the
volume transition in ionic gels~\cite{1, 2} and vesicles~\cite{6, 8}.  The unique feature
in the balloon system is that it is easy to see and
convenient for demonstration.

Though the phenomenon is known to many people, it has not been studied systematically in terms of phase transition. The phenomenon has been studied in the context of mechanical instabilities in non-linear elasticity both experimentally~\cite{17, Goncalves2008} and theoretically~\cite{9, Gent2005, Hill1999}, but it
has not been studied in the context of phase transition.
Especially, the phase diagram of the system has not been drawn.

In this paper, we shall discuss the inflation of a balloon with reference to gas-liquid transition of simple liquids.
We will draw the phase diagram for the inflation behavior (uniform or biphasic), and discuss how it depends on the radius and thickness of the balloon, and also on the material characteristics of the rubber.

To fix the problem, we consider a cylindrical balloon made of thin uniform rubber membrane with thickness $H_0$. The balloon has radius $R_0$ and length $L_0$ in the force free state (see Fig.~\ref{fg1}~(a)).  We assume that the volume $V$ in the balloon is controlled during the inflation. Such control can be achieved by injecting the balloon with incompressible liquids (e.g., water). We also assume that there is a stretching force $F$ acting at both ends of the
balloon (see Fig.~\ref{fg1}~(b)).  Our question is how the equilibrium state of the balloon changes when $V$ and $F$ are changed.

To discuss the equilibrium state, we consider the elastic energy of the rubber forming the balloon.
Since the rubber forming the balloon is thin, the bending moment induced in the rubber is negligible. Therefore, each part of the balloon can be regarded as stretched along three orthogonal directions, by factor  $\lambda_L$ along the axial direction of the cylinder, and  by factor  $\lambda_R$ along the circumferential direction and by factor
$\lambda_H$ along the direction normal to the surface.
Due to the incompressibity of the rubber,
$\lambda_H $ is written as
$\lambda_H =1 / (\lambda_R \lambda_L) $.

Let $g(\lambda_R, \lambda_L, \lambda_H)$ be the elastic energy of the rubber for its deformation per unit volume.  The form of $g(\lambda_R, \lambda_L, \lambda_H)$
has been stuied in many literatures. A widely used form of $g(\lambda_R, \lambda_L, \lambda_H)$ is Neo-Hookian, which can be written as
$g(\lambda_R, \lambda_L, \lambda_H)=GJ/2$, where $G$ is the shear modulus of the rubber and $J$ is defined by
\begin{eqnarray}
\label{def_J}
J=\lambda_{R}^{2}+\lambda_{L}^{2}+\lambda_{H}^{2} - 3.
\end{eqnarray}
The Neo-Hookian model, however, cannot describe the rubber at large deformation; usually rubber cannot be stretched beyond certain limit, and this behavior is described by the Gent model\cite{15}
\begin{eqnarray}
\label{freenergyo}
g(\lambda_{R}, \lambda_{L})
       =- \frac{GJ_{m}}{2}\ln \left(1-\frac{J}{J_{m}} \right ),
\end{eqnarray}
where $J_{m}$ is a material parameter representing the limiting value of $J$.
The Gent model reduces to the Neo-Hookean model when $J$ is small, but deviates strongly from the Neo-Hookean model
when $J$ is large.

For the balloon with fixed volume $V$ subjected to the force $F$, the total free energy is written as
\begin{eqnarray}
   \label{freenergy}
     G_{tot}(\lambda_{R}, \lambda_{L})=2\pi R_{0}H_{0}L_{0} g(\lambda_{R}, \lambda_{L}) - F L.
\end{eqnarray}
The equilibrium state is given by $\lambda_{R}$ and $\lambda_{L}$ which minimize eq.~(\ref{freenergy}) subject to the constraint
$ V=\lambda_{R}^2 \lambda_{L} V_0$, where
$V_0=\pi R_0^2 L_0$ is the initial volume of the balloon.  As a result, the free energy
$G_{tot}$ is obtained as a function of $V$ and $F$.
The pressure $\Delta P$ needed to blow up the balloon is given by
\begin{eqnarray}
\label{pressure}
     \Delta P = \frac{ \partial G_{tot} }{\partial V}.
\end{eqnarray}
Notice that $\Delta P$ denotes the difference in the pressure between inside and outside of the inflating balloon.

\begin{figure}[h]
\begin{center}
\includegraphics{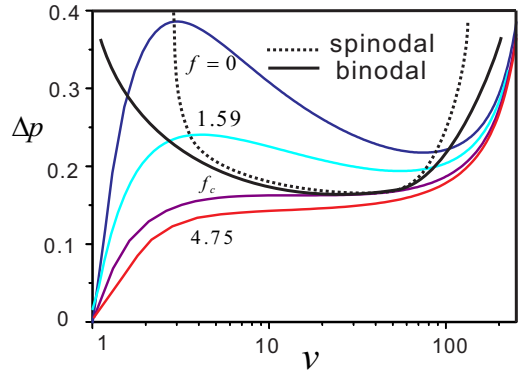}
\caption{
 Pressure difference between inside and outside of balloon $\Delta p$ is plotted against the reduced volume $v$ of the balloon for various uniaxial forces $f$. $f_{c}=3.49$ is a specified value where there exists one critical point on $\Delta p-v$ curve satisfying $d\Delta p/dv=0, d^{2}\Delta p/d^{2}v=0$.
}
\label{fg2}
\end{center}
\end{figure}

The pressure $\Delta P$ obtained in this way is
shown in Fig.~\ref{fg2}, where the pressure $\Delta P$, volume $V$ and the stretching force $F$ are shown in a dimensionless form: $\Delta p = \Delta P R_{0}/2GH_{0}$, $v=V/V_0$, and $f=F/2\pi R_{0}H_{0}G$.

For small $f$, $\Delta p$ is a non-monotonous function of $v$:
$\Delta p $ first increases with $v$ and then decreases and then
increases again. The decrease of $\Delta p$ is due to the non-linear elasticity of the rubber: Neo-Hookean rubber
softens as stretched. The second increase of $\Delta p$
is due to the finite extensibility of the rubber: the stress in the rubber diverges when the rubber is extended up to the
limit represented by $J_m$.

\begin{figure}[h]
\begin{center}
\includegraphics{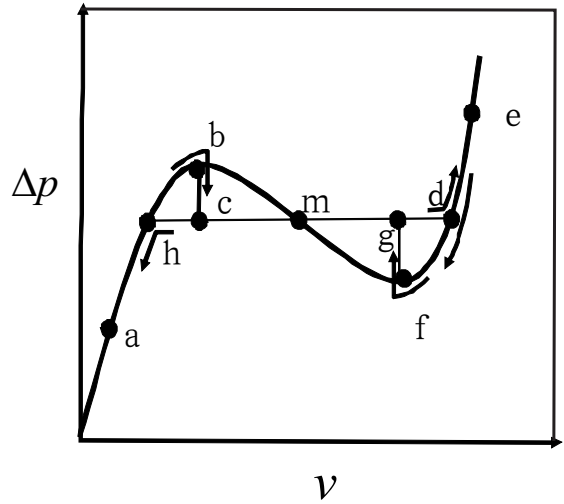}
\caption{Illustration of hysteresis behavior during inflation.}
\label{hysteresis}
\end{center}
\end{figure}

When $\Delta p$ becomes a non-monotonous function of $v$,
consideration is needed to deduce the actual behavior of pressure when $v$ is changed. This is explained in Fig.~\ref{hysteresis}.
Notice that the system is unstable in the region
\emph{bmf} in Fig.~\ref{hysteresis}, where
$\Delta p$ decreases with the increase of $v$.
 In such region, if certain part of the cylindrical balloon is inflated more than neighboring parts, the pressure there becomes less than the neighboring parts, and the inflated part sucks more liquid, and inflates more. Therefore, the state \emph{bmf} is unstable. The points \emph{b} and \emph{f} represent the
limit of the stability and correspond
to the spinodal points in the gas-liquid transition.
The pressure $\Delta p_1$ and $\Delta p_2$ at these points are functions of $f$.

If the volume of the balloon is brought into the unstable region,
the balloon decomposes into strongly inflated
part and weakly inflated part, and becomes at equilibrium  in a biphasic state where the strongly inflated part and the weakly inflated part coexist with each other at a certain pressure $\Delta p_{co}$. The equilibrium state is determined by the condition that the free energy of the whole system is minimum. It can be shown that the pressure $\Delta p_{co}$ is given
by the Maxwell construction, i.e., by finding a horizontal line $\Delta p = \Delta p_{co}$ for which the areas
enclosed by the $\Delta p-v$ curve above and below the line
are equal to each other.

Except for the region \emph{bmf}, the system is stable for small perturbation of $\lambda$ in radial
and axial directions.  Therefore when the volume
$v$ of the balloon is
changed continuously, the pressure $\Delta p$ is
changed as follows (see Fig.~\ref{hysteresis}).
When the volume is increased from the initial state, represented by \emph{a} in Fig.~\ref{hysteresis}, the pressure first increases monotonously along the line \emph{a} $\to $ \emph{b}. At point \emph{b}, phase separation starts (the system decomposes into strongly inflated part and weakly inflated part)
, and the pressure decreases along the line \emph{b} $\to $ \emph{c}. Further increase of the volume is achieved by the
increase of the strongly inflated part (i.e., by the motion of the boundary between the strongly inflated part and weakly inflated part), and the pressure remains constant (\emph{c} $\to $ \emph{d}). At point \emph{d}, the whole balloon becomes strongly inflated. For further increase of the volume, the pressure increases along the line \emph{d} $\to $ \emph{e}. By repeating the same argument, one can show that when the volume of the balloon is decreased, the pressure changes along the line
\emph{e} $\to $ \emph{f} $\to $ \emph{g} $\to $ \emph{h} $\to $ \emph{a}. The points \emph{d} and \emph{h} represent the end points of the biphasic region, and correspond to the binodal points in the gas-liquid transition.

\begin{figure}[h]
\begin{center}
\includegraphics{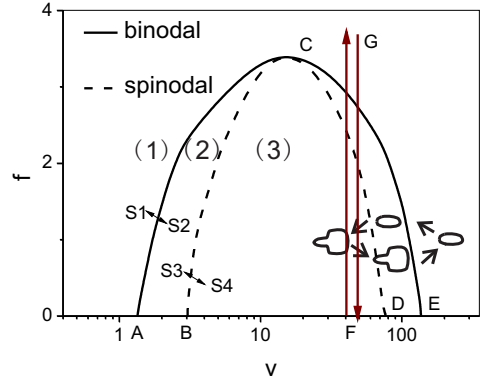}
\caption{
Phase diagram in $f-v$ space for balloon inflation. In region (1), the system must be uniform. In region (3), the system must be biphasic. In region (2), the system can be either uniform or biphasic.
}
\label{fg3}
\end{center}
\end{figure}

The volume at the spinodal points \emph{b} and \emph{f}, and at the binodal points \emph{d} and \emph{h} are functions of the applied force $f$. When $f$ is changed, they give a trajectory in the $\Delta p-v$ plane (see Fig.~\ref{fg3}). The trajectory of the binodal points forms the binodal line (denoted by the solid line), and the trajectory of the spinodal points forms the spinodal line (denoted by the dashed line).
Above the binodal line, only a
uniform state is possible. Below the spinodal line
(in the hatched region), the uniform state is unstable and
the system has to take a biphasic state. In the other region, the system can take either uniform or biphasic state.

The phase diagram (Fig.~\ref{fg3}) indicates where the transition between the uniform state and the biphasic state takes place.
Since thermal energy is negligible in the present macroscopic
system, the transition takes place only when the system becomes unstable for small perturbation in the configuration. Since the biphasic state is stable for small perturbation,
discontinuous transition can take place only when a uniform system is brought into the unstable region by crossing
the spinodal line.
For example, discontinuous transition takes place if the system is in the uniform state at $\rm{S}_3$ in Fig.~\ref{fg3},
and the parameter is changed from $\rm{S}_3$ to $\rm{S}_4$.  On the other hand, discontinuous
phase transition does not take place if the parameter is changed from $\rm{S}_4$ to $\rm{S}_3$ since the system remains
in the biphasic state for this parameter change.

No discontinuous transition takes place at crossing the binodal line. In the crossing of $\rm{S}_1$ to $\rm{S}_2$
in Fig.~\ref{fg3}, the system remains uniform. In the crossing of $\rm{S}_{2}$ $\to $ $\rm{S}_{1}$, the state of the system may change from biphasic to uniform if the system is in a biphasic state at $\rm{S}_2$. However, this transition is continuous: the transition is a result of the
disappearance of the strongly inflated part, and there is no
discontinuous change in the pressure.

\begin{figure}[h]
\begin{center}
\includegraphics{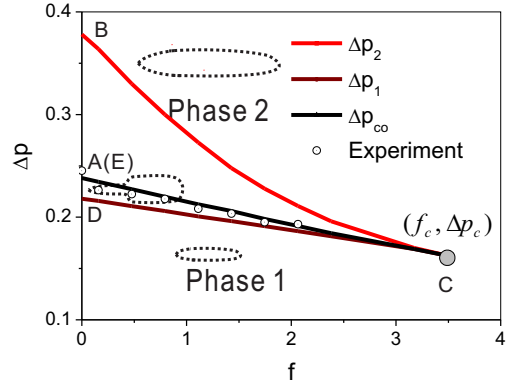}
\caption{
Phase diagram in $\Delta p-f$ space for balloon inflation. Dots are experimental data from Ref.~\cite{17}.
}
\label{fg5}
\end{center}
\end{figure}

Fig.~\ref{fg5} shows the spinodal line and the binodal line in the pressure-force ( $\Delta p - f$ ) space.
The curves BC and DC correspond to the spinodal line BC and DC in
Fig.~\ref{fg3}. A uniform state becomes unstable if it crosses the line BC from below to above, or if it crosses the line DC from above to below. The curve A(E)C corresponds to the binodal line: if the system is in the biphasic state, the pressure $\Delta p$ and the force $f$ must be on this line.

Such phase diagrams are useful in predicting the behavior of the balloon when the external parameters $v$ and $f$ are changed. As an example, let us consider the stretching of a balloon of constant volume
(i.e., a balloon with closed ends). This parameter change is represented by the arrow F $\leftrightarrow $ G in Fig.~\ref{fg3}.

\begin{figure}[h]
\begin{center}
\includegraphics{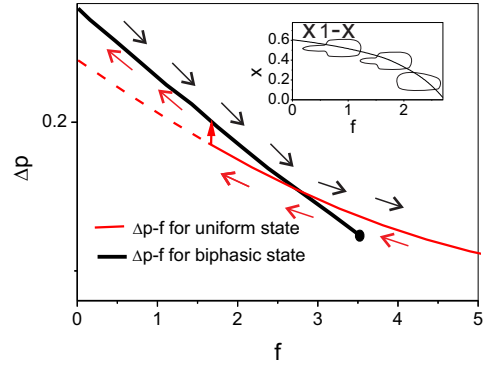}
\caption{
Inflation behavior of closed balloon under stretching. The red curve corresponds to uniform balloon inflation with constant volume $v=55$.
}
\label{fgo}
\end{center}
\end{figure}

At point F, the balloon is in the biphasic state. When the balloon is stretched, it remains in the biphasic state, and  the pressure $\Delta p$ follows the binodal line. The biphasic state ends when the line FG crosses the binodal line in Fig.~\ref{fg3}. Above the binodal line, the balloon is uniformly stretched and the pressure follows the curve of a uniform state. This is represented by the curve in
Fig.~\ref{fgo}.

In the reverse parameter change G $\rightarrow $ F,
the balloon remains in the uniform state if the force $f$
is larger than the value given by the spinodal line DC in Fig.~\ref{fg3}. When the force becomes less than the value of the spinodal line, the uniform state cannot be stable, and
the system becomes biphasic. At this point,
there will be a discontinuous change in the pressure $\Delta p$ as it is indicated in Fig.~\ref{fg2}. Such discontinuous transition for closed balloon has not been
predicted as far as we know.

\begin{figure}[h]
\begin{center}
\includegraphics{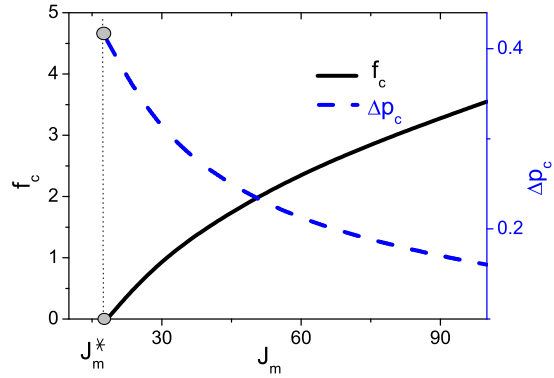}
\caption{
Relation between critical value $(f_{c}, \Delta p_{c})$ and $J_{m}$. $J_{m}^{\ast}=17.2$ is lowermost value for phase coexistence under stretching force.
}
\label{fgy}
\end{center}
\end{figure}

For the discontinuous transitions to be observed, the
pressure $\Delta p$ must be below the critical pressure $\Delta p_c$ and the force $f$ must be below the critical force $f_c$. $\Delta p_c$ and $f_c$ are obtained by the critical condition:
\begin{eqnarray}
\label{pressure}
     \frac {\partial \Delta p}{\partial v}
      = \frac {\partial^2 \Delta p}{\partial^2 v} =0,
\end{eqnarray}
and $\Delta p_c$ and $f_c$ depend on the material parameter of the rubber.  In the Gent model, $\Delta p_c$ and $f_c$ depend only on $J_m$, the stretchability parameter. They are shown in Fig.~\ref{fgy}. With increase of $J_{m}$,
$f_c$ decreases, and becomes zero at $J_m=J_{m}^{\ast}=17.2$. There is no phase coexistence or transition if $J_{m}$ of the material is less than $J_{m}^{\ast}$ no matter how one changes $\Delta p$ or $f$ if keeping force as a stretching one.

We have studied the inflation of a cylindrical rubber balloon under uniaxial stretching force. Our analysis
indicates that the inflation of rubber balloon can be discussed by a theoretical framework which is very similar to gas-liquid transition. The rubber balloon can have two states, a uniform (single phase) state, and biphasic state. The transition between these states can take place continuously or discontinuously, and phase diagram can be drawn in the parameter space of volume $V$, pressure $\Delta P$, and force $F$.
It is predicted that there is a critical point above which there is no discontinuous transition, and that the discontinuous transition is discussed for materials with different stretchability.

The analysis shown here is for cylindrical balloon, but similar analysis can be done for other shaped balloons. Also, actual process of the discontinuous transition can be studied in detail such as where does the nucleation starts and how much extra work is needed to create a stable nucleus of inflated part. These will be discussed in future works.

\begin{acknowledgments}
This work was supported by NSFC under grant No.31128004, 91027045, 10834014, 11175250. FL.M thanks the support of the National Basic Research Program of China (973 program, No.2013CB932800).
\end{acknowledgments}

\providecommand*\mcitethebibliography{\thebibliography}
\csname @ifundefined\endcsname{endmcitethebibliography}
  {\let\endmcitethebibliography\endthebibliography}{}

\end{document}